\documentclass{aa}
\usepackage{txfonts}
\usepackage[latin1]{inputenc}
\usepackage[english]{babel}
\usepackage{xspace}
\usepackage[dvips]{graphicx}
\usepackage{longtable}

\newcommand{\acronym}[1]{\uppercase{#1}\xspace}
\newcommand{\sn}{SN\xspace}
\newcommand{\sne}{SNe\xspace}
\newcommand{\snia}{SN~Ia\xspace}
\newcommand{\sdss}{\acronym{sdss}}
\newcommand{\snls}{\acronym{snls}}
\newcommand{\essence}{\acronym{essence}}

\newcommand{\saltii}{SALT2\xspace}
\newcommand{\sed}{\acronym{sed}}

\newcommand{\jkt}{\acronym{jkt}}
\newcommand{\site}{SITe2\xspace}
\newcommand{\inttel}{\acronym{int}}
\newcommand{\nottel}{\acronym{not}}
\newcommand{\jkttel}{\jkt}
\newcommand{\alfosc}{\acronym{alfosc}}

\newcommand{\isis}{\acronym{isis}}
\newcommand{\wfs}{\acronym{wfs}}
\newcommand{\wfc}{\acronym{wfc}}
\newcommand{\ccd}{\acronym{ccd}}
\newcommand{\fwhm}{\acronym{fwhm}}
\newcommand{\iraf}{\acronym{iraf}}
\newcommand{\sex}{SExtractor\xspace}

\newcommand{\psf}{\acronym{psf}}
\newcommand{\ab}{\acronym{ab}}

\newcommand{\ie}{\mbox{i.e.}\xspace}      
\newcommand{\eg}{\mbox{e.g.}\xspace}
\newcommand{\etal}{\mbox{et al.}}

\newcommand{\supernovaname}[2]{{#1}{#2}\xspace}
\newcommand{\sndr}{\supernovaname{1999}{dr}}
\newcommand{\sndu}{\supernovaname{1999}{du}}
\newcommand{\sndv}{\supernovaname{1999}{dv}}
\newcommand{\sndx}{\supernovaname{1999}{dx}}
\newcommand{\sndy}{\supernovaname{1999}{dy}}

\newcommand{\secref}[1]{Sect.~\ref{#1}\xspace}

\newcommand{\eqref}[1]{(\ref{#1})}

\begin{document}

\title{Light curves of five type Ia supernovae at intermediate redshift%
\thanks{%
  Based on joint obser\-vations made through the Isaac Newton Groups'
  Wide Field Camera Survey Programme with the Isaac Newton Telescope
  and the Jacobus Kapteyn Telescope operated on the island of La Palma
  by the Isaac Newton Group in the Spanish Observatorio del Roque de
  los Muchachos of the Instituto de Astrofisica de Canarias.
  Part of the data presented here have also been taken using ALFOSC,
  which is owned by the Instituto de Astrofisica de Andalucia (IAA)
  and operated at the Nordic Optical Telescope under agreement between
  IAA and the NBIfAFG of the Astronomical Observatory of Copenhagen.
%
%  Parts of the analysis presented here made use of the Perl Data
%  Language (PDL). PDL has been developed by K. Glazebrook, J.
%  Brinchmann, J. Cerney, C. DeForest, D. Hunt, T. Jenness, T. Luka, R.
%  Schwebel, and C. Soeller and can be obtained from
%  \texttt{http://pdl.perl.org}. PDL provides a high-level numerical
%  functionality for the Perl scripting language (Glazebrook \&
%  Economou, 1997).
}}

\author{
  R.~Amanullah\inst{1,2}
  \and V.~Stanishev\inst{1}
  \and A.~Goobar\inst{1}
  \and K.~Schahmaneche\inst{3}
  \and P.~Astier\inst{3}
  \and C.~Balland\inst{3,4}
  \and R.~S.~Ellis\inst{5,6}
  \and S.~Fabbro\inst{7}
  \and D.~Hardin\inst{3}
  \and I.~M.~Hook\inst{8}
  \and M.~J.~Irwin\inst{5}
  \and R.~G.~McMahon\inst{5}
  \and J.~M.~Mendez\inst{9,10}
  \and M.~Mouchet\inst{11,12}
  \and R.~Pain\inst{3}
  \and P.~Ruiz-Lapuente\inst{9}
% \and G.~Sainton\inst{2}
  \and N.~A.~Walton\inst{5}
}

\offprints{R. Amanullah, \email{rahman@physto.se}}
\institute{Physics Department, Stockholm University,
  AlbaNova University Centre, 106~91 Stockholm, Sweden
  \and\emph{Current address:}
    UC Berkeley, Space Sciences Laboratory, 7 Gauss Way, Berkeley, CA,
    $94720-7450$, USA
  \and LPNHE, CNRS-IN2P3 and Universities of Paris 6 \& 7, 75252
    Paris, France
%  \and IAS, UMR 8617 CNRS and Universit\'e Paris-Sud 11, 91405 Orsay,
%  France
  \and Univ. Paris-Sud, Orsay, F-91405, France
  \and Institute of Astronomy, University of Cambridge, Madingley
    Road, Cambridge, CB3 0HA, UK
  \and California Institute of Technology, Pasadena, CA 91125, USA
  \and CENTRA-Centro M. de Astrofisica and Department of Physics, IST,
    Lisbon, Portugal
  \and Astrophysics, Denys Wilkinson Building, Keble Road, OX1 3RH,
    Oxford, UK
  \and Department of Astronomy, University of Barcelona, 08028, Barcelona,
    Spain
  \and Isaac Newton Group of Telescopes, Apartado 321, 38700 Santa
    Cruz de La Palma, Spain
  \and Laboratoire APC, University Paris 7, 10 rue Alice Domon et
    L\'eonie Duquet, 75205 Paris Cedex 13, France
  \and LUTH, UMR 8102 CNRS, Observatore de Paris, Section de Meudon,
    92195 Meudon Cedex, France
}

\date{Received 14 November 2007 / Accepted 21 April 2008}

\abstract{}{We present multi-band light curves and % redshift-luminosity
  distances for five type Ia supernovae at intermediate redshifts,
  $0.18<z<0.27$.}{Three telescopes on the Canary Island of La~Palma,
  \inttel, \nottel, and \jkt, were used for discovery and follow-up of
  type Ia supernovae in the $g'$ and $r'$ filters. Supernova fluxes
  were measured by simultaneously fitting a supernova and host galaxy
  model to the data, and then calibrated using star catalogues from
  the Sloan Digital Sky Survey.}{The light curve shape and colour
  corrected peak luminosities are consistent with the expectations of
  a flat $\Lambda$CDM universe at the $1.5\sigma$ level.
  One supernova in the sample, SN\sndr, shows surprisingly large
  reddening, considering both that it is located at a significant
  distance from the core of its host ($\sim4$ times the fitted
  exponential radius) and that the galaxy can be spectroscopically
  classified as early-type with no signs of ongoing star formation.
  }{}


\maketitle
  
%
%________________________________________________________________

\section{Introduction}
Measurements of type Ia supernovae (\snia) have had a dra\-ma\-tic
im\-pact on the field of cosmology, and lead to the discovery of the
acceleration of the Universe
(\cite{1998AJ....116.1009R,1999ApJ...517..565P}). Since then the
number of well measured \snia:s has increased considerably through the
joint effort of several independent research groups. However, the
majority of the search campaigns have either been looking for high
redshift \snia
(\cite{knop03,2006A&A...447...31A,riess07,astro-ph/0701041} and
references therein) where the difference in the relative luminosity
distance between different cosmological models increases, as shown by
Goobar \& Perlmutter (1995), or low redshift \snia
(\cite{1996AJ....112.2438H,1999AJ....117..707R,jha06}) which are
essential for both anchoring the Hubble diagram as well as learning
more about \snia properties. With the exception of the ongoing three
year Sloan Digital Sky Survey \sn campaign initiated in 2005
(\cite{astro-ph/0708.2749,astro-ph/0708.2750}), the redshift region
around $z\sim0.2$ in the Hubble diagram has not been heavily explored.

This missing link in the \snia Hubble diagram is of great importance
as it corresponds to the epoch of dark energy domination. Thus it
presents interesting opportunities for testing for a possible time
evolution of the dark energy density with data-sets probably less
plagued by systematic uncertainties. Effects such as \snia brightness
evolution or gravitational lensing are expected to be much less
than at higher redshifts.

The current best estimates of cosmological parameters from \snls
(\cite{2006A&A...447...31A}) and \essence (\cite{astro-ph/0701041})
are approaching the systematic uncertainties for these projects. This
further increases the need of a better understanding of \snia
properties with the aim of calibrating them to even better precision.

\begin{figure*}[!tb]
  \centering
  \includegraphics[height=\textwidth,angle=-90]{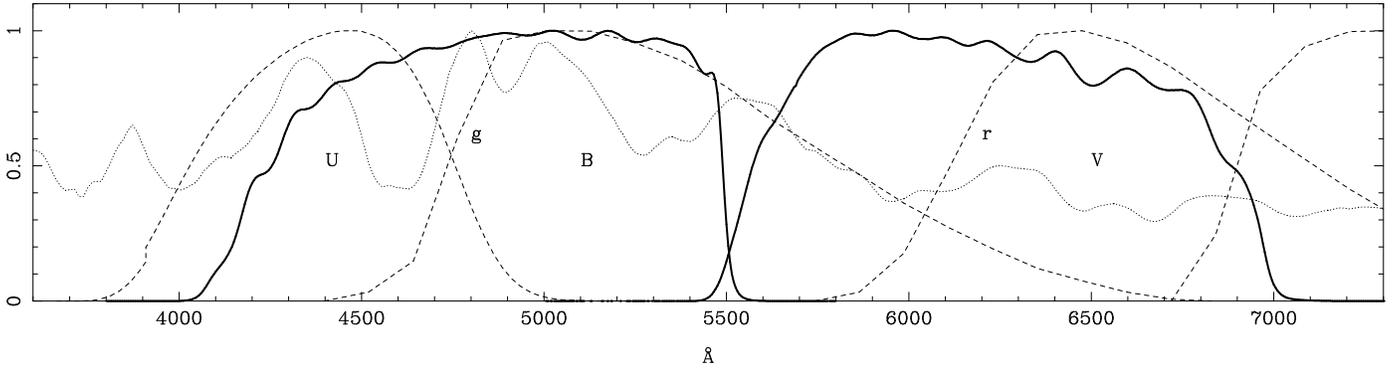}
  \caption{Overlap between rest frame $UBV$ (dashed) and observed,
    effective, \inttel $g'$ and $r'$ (solid) filters for the mean
    redshift, $z=0.22$, of the \sn sample presented here. The dotted
    curve shows the normalised spectral template of a type~Ia
    supernova (\cite{2002PASP..114..803N}) at maximum.
%    The effective $g'$ and $r'$ filter curves consist of the measured
%    filter response, the quantum efficiency of the Wide Field Camera, the
%    expected atmospheric transmission at La~Palma (airmass=1), and the
%    reflectivity of the mirror.
    \label{fig:intzfilters}}
\end{figure*}

Also, for these redshifts the rest frame ultraviolet region of the
\snia spectrum has been redshifted into the visible wavelength area,
illustrated in Figure~\ref{fig:intzfilters}, where the detector
efficiency is higher and the atmosphere is transparent and less
variable.
This is a part of the type~Ia spectrum that contains a large fraction
of the total flux, and it is therefore very important to understand
the intrinsic UV-properties for interpreting high redshift data.

Furthermore, the UV part of the spectrum is of particular interest as
it has been suggested that a redshift evolution of the \snia
properties due to differences in the progenitor metallicity, if
present, may be detectable in this wavelength region
(\cite{1998ApJ...495..617H,2000ApJ...530..966L}). Intermediate
redshift \sne are bright enough to allow high precision spectroscopy
(\cite{2008ApJ...674...51E}), and a large number of well observed
\snia at these redshifts may very well turn out to be essential for
understanding rest frame $U$-band properties.

To address these issues, a pilot campaign for photometric and
spectroscopic follow-up of intermediate redshift \snia was launched in
1999. Here we present photometric follow-up of five \snia from the
1999 campaign. The spectra of these \sne have already been published
in Balland~\etal~(2006).

In \secref{sec:searchfollow} we describe the search and the
photometric follow-up. \secref{sec:redphot} covers the reduction
and photometric measurements, followed by a description of the light
curve analysis and possible systematic errors
in~\secref{sec:analysis}. In \secref{sec:discussion} the results are
summarised and discussed.

%
%________________________________________________________________

\section{Search and Follow-up Observations}\label{sec:searchfollow}
The search campaign was carried out during autumn 1999 as a piggy-back
project of the Wide Field Survey (\wfs) (\cite{2001NewAR..45...97M})
at the Isaac Newton Telescope (\inttel). The \wfs obtained multi-colour
data in the broad band, $u'$, $g'$, $r'$, $i'$, $z'$ filter set
during 1998--2003, covering 200~square degrees with a depth of
$r'\approx24$ and $g'\approx25$. The Wide Field Camera (\wfc), with
four 4k$\times$2k {\ccd}s and a total field of view of $0.27$~square
degrees, was used and the $g'$ and $r'$ data was obtained repeatedly
with a cadence particularly suitable for finding \sne\ around maximum
(\cite{1995ApJ...440L..41P}, 1997).

The \sne presented here were discovered by seeing-matching and
comparing two \wfs data sets in the $g'$-band. The first set was
observed in mid August 1999 and the second roughly one month later.
The August data consisted of 600~s exposures, while two 240~s frames
were obtained for the same fields in September~1999. Altogether, three
subtractions were made by both considering the September discovery
images individually, and by using their coadded sum. Candidates were
then selected by requiring detection within 2~pixels on all three
frames in order to efficiently reject cosmic rays and fast-moving
objects. In addition to this, a flux increase within one \fwhm seeing
radius of at least 15\ts\% was used as a search criteria. The
candidates were re-observed 1--2 days later and rejected unless the
above criteria were still fulfilled. This excluded possible
slow-moving objects from the candidate sample. The whole search
resulted in a total of 15~\sn candidates.

Spectroscopy of the candidates was obtained using the red channel of
the \isis instrument on the 4.2~m William Herschel Telescope.
Additional spectroscopy was obtained for two of the candidates, (\sndu
and \sndv), using the 2.5~m Nordic Optical Telescope (\nottel) three
weeks after discovery. Ten of the \sne, shown in
Table~\ref{tb:intzallsne} with their approximate discovery magnitude
and spectroscopic redshifts, were confirmed as \snia
(\cite{2006A&A...445..387B}).
\begin{table}[tb]
  \centering
  \caption{The ten \snia discovered in the autumn 1999 campaign.
    \label{tb:intzallsne}}
  \begin{tabular}{c@{\hspace{1ex}} l l r r l c}
    \hline
    &
    \multicolumn{1}{c}{\small\sn}  &
    \multicolumn{1}{c}{\small$\alpha\ (2000)$} &
    \multicolumn{1}{c}{\small$\delta\ (2000)$} &
    \multicolumn{1}{c}{\small$g'$} &
    \multicolumn{1}{c}{\small$z$}\\
    \hline
    $\rightarrow$ &
	1999dr & 23 00 17.56 & $-$00 05 12.5 & 22.1 & 0.178\\% & Ia\\
%     & 1999ds & 23 53 51.63 & $+$00 09 22.9 & 22.4 & 0.350\\% & II?\\
      & 1999dt & 00 45 42.29 & $+$00 03 22.2 & 23.5 & 0.437\\% & Ia\\
    $\rightarrow$ &
	1999du & 01 07 05.94 & $-$00 07 53.8 & 22.8 & 0.260\\% & Ia\\
    $\rightarrow$ &
        1999dv & 01 08 58.96 & $+$00 00 24.8 & 21.8 & 0.186\\% & Ia\\
      & 1999dw & 01 22 52.80 & $-$00 16 20.8 & 24.1 & 0.460\\% & Ia?\\
    $\rightarrow$ &
	1999dx & 01 33 59.45 & $+$00 04 15.3 & 22.2 & 0.269\\% & Ia\\
    $\rightarrow$ &
	1999dy & 01 35 49.53 & $+$00 08 38.3 & 21.7 & 0.215\\% & Ia\\
      & 1999dz & 01 37 03.24 & $+$00 01 57.9 & 23.4 & 0.486\\% & Ia\\
      & 1999ea & 01 47 26.09 & $-$00 02 07.2 & 23.3 & 0.397\\% & Ia\\
      & 1999gx & 00 34 15.47 & $+$00 04 26.2 & 23.3 & 0.493\\% & Ia\\
    \hline
  \end{tabular}
\end{table}
The redshifts could be determined from host galaxy features for all
\sne and the precision of these measurements is typically 0.001.

Five of the \snia had redshifts below $z<0.3$, marked with arrows in
Table~\ref{tb:intzallsne}, and were all discovered before or around
maximum. For these, photometric follow-up in the $g'$ and $r'$ bands
was carried out using the \alfosc instrument on the 2.5~m \nottel and
the \site \ccd camera on the 1.0~m Jacobus Kapteyn Te\-lescope (\jkt)
in addition to \inttel. This filter combination corresponds
approximately to rest frame $UBV$ for $z\sim0.2$. For the
remaining SNe ($z>0.3$), we do not have any photometric follow-up.
%\ad{The remaining 5 SNe were too distant ($z>0.3$) to allow
%  accurate follow-up well past maximum with the available telescopes.}

\sn-free reference images of the host galaxies were obtained
approximately one year after discovery with \inttel in both $g'$ and
$r'$.

\section{Reduction and Photometry}\label{sec:redphot}
The \inttel images were reduced using the \wfs pipeline
(\cite{2001NewAR..45..105I}). For the \nottel and \jkttel images,
standard reduction including bias subtraction and flat fielding, was
carried out using the \iraf%
\footnote{\iraf is distributed by the National Optical Astronomy
  Observatories, which are operated by the Association of Universities
  for Research in Astronomy, Inc., under cooperative agreement with
  the National Science Foundation.}
software. A spatially varying multiplicative factor was still present
for the \jkttel images after flat fielding, which possibly could
originate from small filter movements during the night. The effect was
corrected for by fitting a smooth surface to each image. Some
images showed signs of bad telescope tracking and were rejected.

%
%________________________________________________________________

\subsection{Photometry}\label{sec:photometry}
The photometry technique applied for this data set is based on the
same method (\cite{FabbroThesis}) used for the latest \snls data release
(\cite{2006A&A...447...31A}).

For each exposure covering a given \sn, the \sex package
(\cite{1996A&AS..117..393B}) was first used to create object
catalogues. These were used to fit geometric transformations to the
best seeing \emph{photometric reference} image of the sample and all
other images were then re-sampled to this frame. In order to properly
compare images obtained during different seeing conditions we fitted
convolution kernels, modelled by a linear decomposition of Gaussian
and polynomial basis functions
(\cite{1998ApJ...503..325A,2000A&AS..144..363A}), between the
photometric reference and each of the remaining images for the given
passband. The kernels were fitted by using image patches centred on
objects across the field. The quality of both the geometric
transformations and the fitted kernels was investigated by carrying
out image subtractions and searching for residual artifacts. Such
artifacts were in general absent and had minimal impact on the \sn
photometry.

For a small area centred on an \sn, the \psf is not expected to have
any spatial variation. A time series of such patches, $I_i$, can be
described by the following model
\begin{displaymath}
  I_i(x,y) = f_i\cdot\left[K_i\otimes\mathrm{\psf}\right](x -
    x_0, y - y_0 ) + \left[K_i\otimes G\right](x,y)\, + S_i\, .
\end{displaymath}
Here, $I_i(x,y)$ is the value in pixel $(x,y)$ of image $i$, $f_i$ is
the \sn flux, \psf is the point spread function for the photometric
reference, $K_i$ is the kernel between the reference and image $i$,
$\otimes$ is the convolution operator, $(x_0,y_0)$ is the
\sn-position, $G$ is a time independent galaxy model and $S_i$ is the
local sky background.

The \psf was obtained from the field stars on the photometric
reference image. An initial guess of the \sn-position was found by
stacking all images that were believed to contain \sn-light and
subtracting the stack of \sn-free reference images from this. The
parameters of the model, $f_i$, $x_0$, $y_0$, $G$ and $S_i$ were then
simultaneously fitted to the whole patch set by minimising
\begin{displaymath}
  \chi^2 = \sum_i^N\sum_{x,y}^{k,n} W_i(x,y)\cdot%
    \left[D_i(x,y) - I_i(x,y)\right]^2\, ,
\end{displaymath}
where $D_i(x,y)$ is the data value in pixel $(x,y)$ on image patch $i$
and $W_i(x,y)$ is the weight. The Poisson and read-out noise as well
as kernel and \psf uncertainties were propagated to $W_i$. The sums
were carried out over the total number of image patches, $N$, and the
patch dimensions, $k\times n$. In this work, we used a non-analytic
host galaxy model with one parameter for each pixel. This means that
the model is degenerate with the \sn and in order to break this
degeneracy we kept $f_i=0$ fixed for the \sn-free \inttel images
obtained in 2000. The model is also degenerate with the local sky
background, $S_i$, that needs to be fixed to zero on one image. The
total number of fitting parameters will then add up to: $N-R$ \sn
fluxes where $R$ is the number of \sn-free images, $k\cdot n$ for the
time-independent galaxy model, $N-1$ sky values, and 2 for the \sn
position, \ie $1 + 2N - R + k\cdot n$.

\begin{figure*}[!tbp]
  \centering
  \includegraphics[width=\textwidth]{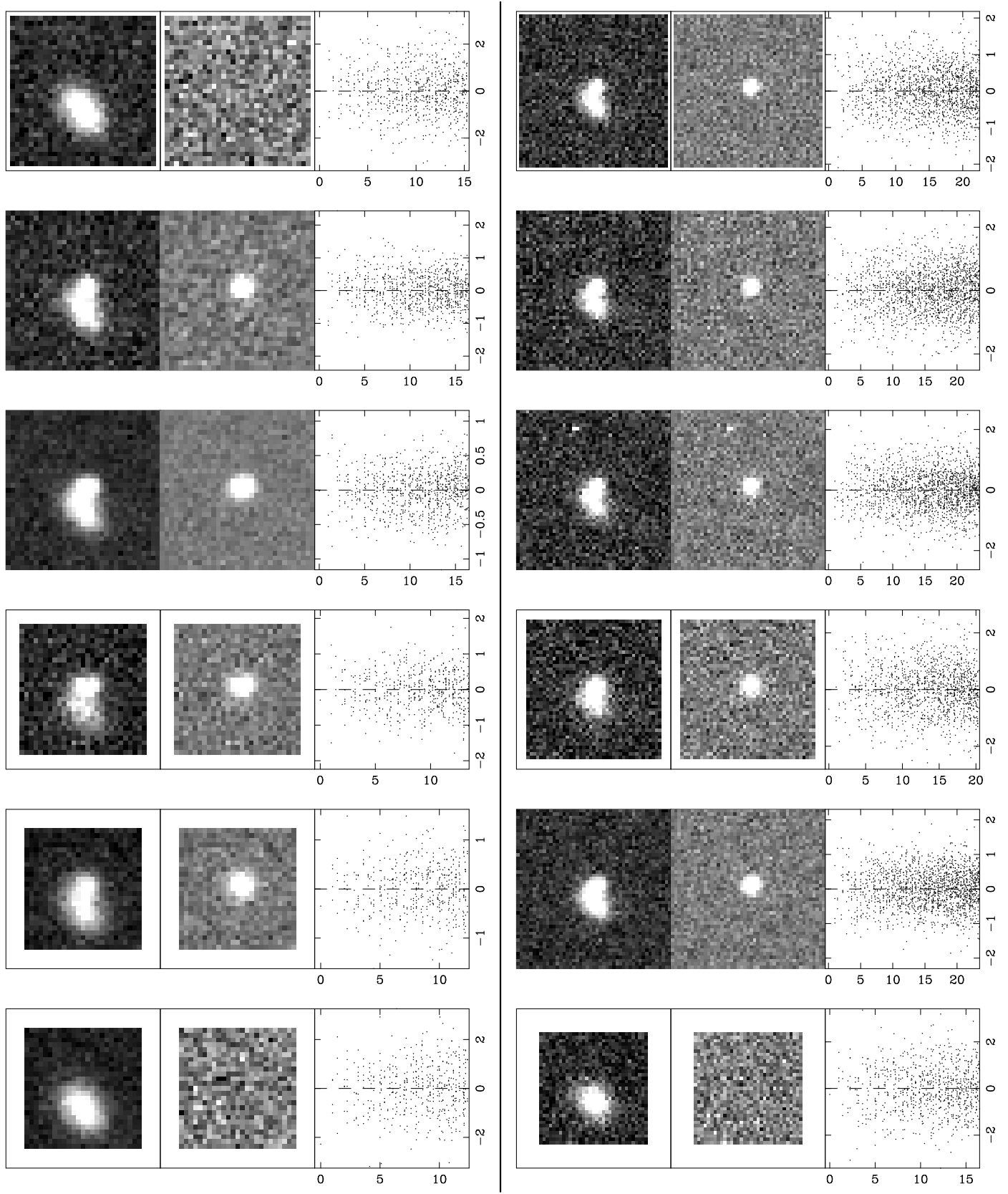}
  \caption{A sample of patches for increasing epochs from the $g'$
    (left column) and $r'$-band (right column) light curve builds of
    \sndu.
   Each triplet represent first the untouched data patch, the
   galaxy model subtracted data patch, and a profile plot where the
   full galaxy $+$ \psf model has been subtracted from the data
   patch. The profile plot shows the deviation from zero in standard
   deviations vs the distance from the \sn position in pixels. The
   last row of the plot shows the \sn-free epoch, obtained
   approximately one year after discovery. The different patch sizes
   reflect the differences in image quality. A small patch size
   represents good seeing conditions.%
    \label{fig:1999dupatches}}
\end{figure*}

An example of image, galaxy model residuals, together with the
resulting residuals when the full model has been subtracted, can be
seen in Figure~\ref{fig:1999dupatches} for increasing epochs. Here the
profile residual plots show the deviation from zero in standard
deviations vs the distance from the \sn position in pixels. The last
row of the plot shows the \sn-free epoch, obtained approximately one
year after discovery. Different patch sizes reflect the differences in
image quality where a small patch size represents good seeing
conditions.

% 1999du patch series
%

The fitted \sn-fluxes are presented in
Table~\ref{tb:photometry}\footnote{This table is only available in
the online version of the paper.}, and the corresponding instrumental
magnitudes in the \inttel system can be obtained as
$m_{\mathrm{\inttel}} = -2.5\log_{10}f + ZP_{AB}$, where $f$ and the
$ZP_{AB}$ is the flux and AB zero point from the table. This table
also shows the image quality (I.Q.) of each epoch and is defined as
the FWHM in arc seconds of a point source. \onllongtab{2}{%
  \centering
  \begin{longtable}{l r c r c r@{\ }r r c}
    \caption{Follow-up photometry of the five \sne presented in this
      work.\label{tb:photometry}}\\
    \hline\hline
    \multicolumn{1}{c}{\sn} &
    \multicolumn{1}{c}{MJD} &
    \multicolumn{1}{c}{Tel.} &
    \multicolumn{1}{c}{Exp. time} &
    \multicolumn{1}{c}{Band} &
    \multicolumn{2}{c}{Flux} &
    \multicolumn{1}{c}{$ZP_{AB}$} &
    \multicolumn{1}{c}{I.Q.}\\
    \hline
    \endfirsthead
    \hline
    \multicolumn{1}{c}{\sn} &
    \multicolumn{1}{c}{MJD} &
    \multicolumn{1}{c}{Tel.} &
    \multicolumn{1}{c}{Exp. time} &
    \multicolumn{1}{c}{Band} &
    \multicolumn{2}{c}{Flux} &
    \multicolumn{1}{c}{$ZP_{AB}$} &
    \multicolumn{1}{c}{I.Q.}\\
    \hline
    \endhead
    \hline\multicolumn{7}{r}{\it continues on next page}
    \endfoot
    \hline\hline
    \endlastfoot
    % \\[0.5ex]
    \sndr & $51399.98$ & \inttel & 599 & $g'$ & 1661 & (360) & 31.80 (0.02) & 2.00\\
\sndr & $51422.99$ & \inttel & 638 & $g'$ & 6775 & (366) & 31.80 (0.02) & 1.30\\
\sndr & $51423.99$ & \inttel & 239 & $g'$ & 6285 & (408) & 31.80 (0.02) & 1.20\\
\sndr & $51427.06$ & \inttel & 599 & $g'$ & 4245 & (277) & 31.80 (0.02) & 1.20\\
\sndr & $51430.93$ & \inttel & 599 & $g'$ & 2755 & (238) & 31.80 (0.02) & 1.20\\
\sndr & $51430.93$ & \inttel & 599 & $r'$ & 7771 & (503) & 31.64 (0.02) & 1.00\\
\sndr & $51434.99$ & \jkttel & 1800 & $r'$ & 4082 & (15001) & 31.64 (0.02) & 2.40\\
\sndr & $51435.09$ & \jkttel & 5400 & $g'$ & 1437 & (510) & 31.80 (0.02) & 2.50\\
\sndr & $51435.97$ & \jkttel & 5400 & $g'$ & 1989 & (258) & 31.80 (0.02) & 1.20\\
\sndr & $51436.90$ & \jkttel & 5400 & $r'$ & 6056 & (1412) & 31.64 (0.02) & 2.40\\
\sndr & $51454.89$ & \jkttel & 977 & $r'$ & 2450 & (769) & 31.64 (0.02) & 1.00\\
\sndr & $51454.99$ & \inttel & 2396 & $g'$ & 638 & (127) & 31.80 (0.02) & 1.70\\
\sndr & $51455.07$ & \inttel & 3595 & $r'$ & 2403 & (373) & 31.64 (0.02) & 2.10\\
\sndr & $51455.90$ & \inttel & 599 & $r'$ & 2253 & (361) & 31.64 (0.02) & 1.00\\
\sndr & $51456.02$ & \jkttel & 3600 & $g'$ & 1030 & (246) & 31.80 (0.02) & 1.00\\
\sndr & $51456.90$ & \jkttel & 5400 & $r'$ & 2272 & (598) & 31.64 (0.02) & 1.50\\
\sndr & $51457.89$ & \jkttel & 3600 & $r'$ & 2218 & (655) & 31.64 (0.02) & 1.50\\
\sndr & $51458.90$ & \jkttel & 3600 & $g'$ & 681 & (565) & 31.80 (0.02) & 2.10\\
\hline
\sndu & $51429.16$ & \inttel & 479 & $g'$ & 8018 & (404) & 32.00 (0.02) & 1.00\\
\sndu & $51431.09$ & \inttel & 599 & $g'$ & 11757 & (493) & 32.00 (0.02) & 1.10\\
\sndu & $51431.10$ & \inttel & 599 & $r'$ & 23228 & (1820) & 32.78 (0.02) & 1.10\\
\sndu & $51437.06$ & \jkttel & 5400 & $g'$ & 20052 & (1227) & 32.00 (0.02) & 2.00\\
\sndu & $51438.05$ & \jkttel & 5400 & $r'$ & 40797 & (2884) & 32.78 (0.02) & 1.30\\
\sndu & $51452.92$ & \nottel & 3600 & $g'$ & 15038 & (566) & 32.00 (0.02) & 1.20\\
\sndu & $51452.96$ & \nottel & 3000 & $r'$ & 37554 & (2150) & 32.78 (0.02) & 1.00\\
\sndu & $51454.93$ & \nottel & 3600 & $r'$ & 36857 & (2108) & 32.78 (0.02) & 1.30\\
\sndu & $51454.97$ & \nottel & 3000 & $g'$ & 13266 & (531) & 32.00 (0.02) & 1.50\\
\sndu & $51456.06$ & \inttel & 1798 & $r'$ & 36466 & (2055) & 32.78 (0.02) & 1.20\\
\sndu & $51456.07$ & \inttel & 1798 & $g'$ & 12400 & (478) & 32.00 (0.02) & 1.30\\
\sndu & $51456.98$ & \jkttel & 5400 & $g'$ & 12676 & (747) & 32.00 (0.02) & 1.70\\
\sndu & $51457.05$ & \jkttel & 4400 & $r'$ & 33045 & (2491) & 32.78 (0.02) & 1.40\\
\sndu & $51458.06$ & \inttel & 1198 & $r'$ & 33341 & (2022) & 32.78 (0.02) & 1.10\\
\sndu & $51458.07$ & \inttel & 1198 & $g'$ & 10831 & (453) & 32.00 (0.02) & 1.10\\
\sndu & $51459.10$ & \jkttel & 2216 & $r'$ & 28781 & (4618) & 32.78 (0.02) & 2.50\\
\sndu & $51461.06$ & \inttel & 599 & $r'$ & 27689 & (1962) & 32.78 (0.02) & 1.30\\
\sndu & $51461.07$ & \inttel & 599 & $g'$ & 8301 & (409) & 32.00 (0.02) & 1.40\\
\sndu & $51597.99$ & \nottel & 600 & $g'$ & 13775 & (722) & 32.00 (0.02) & 1.70\\
\hline
\sndv & $51430.09$ & \inttel & 479 & $g'$ & 50185 & (2195) & 33.00 (0.02) & 1.20\\
\sndv & $51431.12$ & \inttel & 599 & $r'$ & 43118 & (1880) & 32.76 (0.02) & 1.00\\
\sndv & $51431.18$ & \inttel & 599 & $g'$ & 57484 & (2280) & 33.00 (0.02) & 1.00\\
\sndv & $51436.07$ & \jkttel & 3600 & $g'$ & 80463 & (3304) & 33.00 (0.02) & 1.10\\
\sndv & $51436.12$ & \jkttel & 3600 & $r'$ & 61249 & (2799) & 32.76 (0.02) & 1.10\\
\sndv & $51453.06$ & \nottel & 2400 & $r'$ & 47881 & (2075) & 32.76 (0.02) & 1.20\\
\sndv & $51454.11$ & \jkttel & 3600 & $r'$ & 48940 & (3096) & 32.76 (0.02) & 1.40\\
\sndv & $51455.02$ & \nottel & 3000 & $g'$ & 30991 & (1563) & 33.00 (0.02) & 1.50\\
\sndv & $51455.07$ & \nottel & 3600 & $r'$ & 42884 & (1837) & 32.76 (0.02) & 1.30\\
\sndv & $51456.10$ & \inttel & 598 & $r'$ & 42163 & (2058) & 32.76 (0.02) & 1.00\\
\sndv & $51457.13$ & \jkttel & 5400 & $g'$ & 25043 & (1941) & 33.00 (0.02) & 1.20\\
\sndv & $51458.08$ & \inttel & 1198 & $r'$ & 37552 & (1674) & 32.76 (0.02) & 1.10\\
\sndv & $51458.10$ & \inttel & 1198 & $g'$ & 23604 & (1313) & 33.00 (0.02) & 1.30\\
\sndv & $51458.97$ & \jkttel & 3600 & $r'$ & 31942 & (4161) & 32.76 (0.02) & 2.30\\
\sndv & $51461.10$ & \inttel & 599 & $r'$ & 32740 & (1666) & 32.76 (0.02) & 1.30\\
\sndv & $51461.11$ & \inttel & 599 & $g'$ & 16803 & (1174) & 33.00 (0.02) & 1.20\\
\hline
\sndx & $51429.21$ & \inttel & 239 & $g'$ & 10353 & (592) & 31.97 (0.02) & 1.00\\
\sndx & $51431.20$ & \inttel & 599 & $g'$ & 11994 & (566) & 31.97 (0.02) & 1.00\\
\sndx & $51437.18$ & \jkttel & 6102 & $g'$ & 10953 & (1035) & 31.97 (0.02) & 2.20\\
\sndx & $51438.13$ & \jkttel & 5400 & $r'$ & 28422 & (2266) & 32.79 (0.02) & 1.30\\
\sndx & $51453.11$ & \nottel & 3600 & $g'$ & 2271 & (296) & 31.97 (0.02) & 1.00\\
\sndx & $51453.25$ & \nottel & 600 & $r'$ & 22849 & (6849) & 32.79 (0.02) & 1.60\\
\sndx & $51455.13$ & \nottel & 3000 & $r'$ & 10323 & (1429) & 32.79 (0.02) & 1.10\\
\sndx & $51455.17$ & \nottel & 3600 & $g'$ & 1925 & (262) & 31.97 (0.02) & 1.30\\
\sndx & $51456.11$ & \jkttel & 1800 & $g'$ & 2595 & (599) & 31.97 (0.02) & 1.00\\
\sndx & $51456.16$ & \inttel & 599 & $r'$ & 12182 & (1537) & 32.79 (0.02) & 1.20\\
\sndx & $51457.05$ & \inttel & 1198 & $g'$ & 1718 & (252) & 31.97 (0.02) & 1.10\\
\sndx & $51457.07$ & \inttel & 1198 & $r'$ & 10063 & (1097) & 32.79 (0.02) & 1.00\\
\sndx & $51457.96$ & \jkttel & 5400 & $r'$ & 10326 & (1844) & 32.79 (0.02) & 1.30\\
\sndx & $51458.04$ & \jkttel & 3600 & $g'$ & 1464 & (474) & 31.97 (0.02) & 1.40\\
\sndx & $51462.12$ & \inttel & 599 & $g'$ & 1280 & (281) & 31.97 (0.02) & 1.50\\
\sndx & $51462.12$ & \inttel & 598 & $r'$ & 8180 & (1544) & 32.79 (0.02) & 1.60\\
\hline
\sndy & $51429.20$ & \inttel & 479 & $g'$ & 19205 & (751) & 31.82 (0.02) & 1.00\\
\sndy & $51431.20$ & \inttel & 599 & $g'$ & 22519 & (818) & 31.82 (0.02) & 0.90\\
\sndy & $51438.18$ & \jkttel & 1800 & $g'$ & 25252 & (1262) & 31.82 (0.02) & 1.30\\
\sndy & $51438.23$ & \jkttel & 2400 & $r'$ & 88645 & (4827) & 33.18 (0.02) & 1.50\\
\sndy & $51453.20$ & \nottel & 2400 & $g'$ & 12378 & (720) & 31.82 (0.02) & 1.70\\
\sndy & $51453.22$ & \nottel & 2400 & $r'$ & 56868 & (2790) & 33.18 (0.02) & 1.40\\
\sndy & $51454.99$ & \jkttel & 1800 & $r'$ & 56156 & (4367) & 33.18 (0.02) & 1.70\\
\sndy & $51455.22$ & \nottel & 3600 & $g'$ & 10136 & (531) & 31.82 (0.02) & 1.60\\
\sndy & $51455.25$ & \nottel & 600 & $r'$ & 50133 & (3401) & 33.18 (0.02) & 1.50\\
\sndy & $51456.15$ & \jkttel & 3600 & $g'$ & 9185 & (657) & 31.82 (0.02) & 1.30\\
\sndy & $51456.18$ & \inttel & 599 & $r'$ & 50905 & (2564) & 33.18 (0.02) & 1.20\\
\sndy & $51458.11$ & \jkttel & 1800 & $r'$ & 48287 & (5086) & 33.18 (0.02) & 1.80\\
\sndy & $51458.16$ & \jkttel & 1800 & $g'$ & 7374 & (1599) & 31.82 (0.02) & 2.60\\
\hline

  \end{longtable}}

When multiple exposures are available for a given night, the measured
\sn fluxes are compared to each other in order to make sure that they
agree within their statistical uncertainty. This was the case for all
nights except for the \sndx $g'$ exposures from 1999-09-08 where,
after a more careful investigation, it was concluded that one of the
exposures was plagued with a cosmic ray very close to the \sn, and the
image was therefore excluded.

\subsection{Subtraction Precision}
In order to test the robustness of the photometric method and check
for systematic errors, the fitted \sn + galaxy model was subtracted
from each data patch. Possible subtraction and galaxy model errors
were explored both by measuring the residual flux within one \fwhm
aperture radius on the subtractions, and by studying how the fitted
\sn-fluxes of any given epoch is changing as more epochs are added to
the fit. These tests did not result in any deviations exceeding the
statistical expectations.

\subsection{Calibration}\label{sec:intzlcfitcalib}
All fitted \sn fluxes from the method described above are by
construction tied to the normalisation of the fitted \psf of the
photometric reference image. In order to calibrate the \sn photometry
it is therefore necessary to determine the zero point that relates
this normalisation to a standard photometric system. For this purpose
we used the field stars in the images as tertiary standard stars. The
instrumental magnitudes of these objects were measured using the exact
same procedure as for the \sne except no host galaxy models were
fitted.

In principle, the magnitudes of the tertiary standard stars could be
determined from observations of standard stars during the nights.
However, since that data was not always available, and in order to
obtain a consistent calibration for all \sne, we decided to use
photometric catalogues of our fields from the Sloan Digital Sky Survey
(\sdss) DR4 (\cite{2006ApJS..162...38A}), and tie our calibration to
their photometric system (\cite{1996AJ....111.1748F}), which for $g'$
and $r'$ is very close to the standard \ab system
(\cite{1983ApJ...266..713O}).
From the \sdss catalogues we chose to use \psf magnitudes
and converted them from \sdss magnitudes (\cite{1999AJ....118.1406L}) to
standard Pogson magnitudes (\cite{1856MNRAS..17...12P}).

Since the \inttel data dominates the sample, and since we only had
\sn-free reference images obtained with this telescope, we
consistently chose to use this as our natural magnitude system.
Systematic effects on \nottel and \jkttel points due to this are
discussed below. Zero points were then determined by simultaneously
fitting the following relation between our instrumental magnitudes,
$m_j$, and the \sdss magnitudes, $m^\mathrm{\sdss}_j$, to \emph{all}
\inttel images for a given filter
\begin{equation}
  \Delta m \equiv m_j - m^\mathrm{\sdss}_j = zp_i +
  c_X\cdot(g_j-r_j)^\mathrm{\sdss}\,.\label{eq:sdsscalibration}
\end{equation}
Here $zp_i$ is the zero point for image $i$, $c_{X}$ is the colour
term for filter $X$, and $(g'_j-r'_j)^\mathrm{\sdss}$ is the \sdss
$g'-r'$ colour of the star $j$. The fit was carried out simultaneously
in two iterations, using a 3$\sigma$ outlier rejection cut.

\begin{figure}[htb]
  \centering
    \resizebox{\hsize}{!}{\includegraphics[angle=-90]{%
      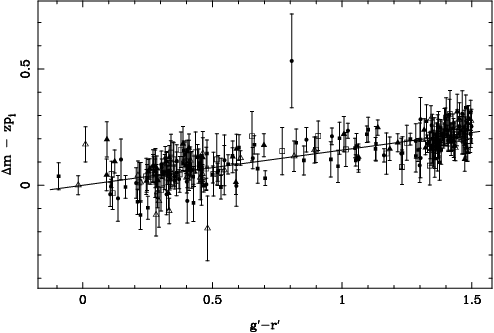}}%
  \caption{Fitted zero point solution in the \inttel $g'$-band.
    \label{fig:intgzpfit}}
\end{figure}

Both the zero points of the individual images and the common colour
terms were obtained this way. The results and the number of stars
used for the fits are presented in Table~\ref{tb:simuzpfit} and the
first row of Table~\ref{tb:colourterms}. The simultaneous fit for
the $g'$ band is also presented Figure~\ref{fig:intgzpfit}, where
$\Delta m$ denotes the difference between the measured instrumental
magnitude and the \sdss magnitude. From this the individually fitted
zero points have been subtracted in order to put all data points on
the same scale and the magnitude differences are then plotted
against the \sdss colours of the stars. Different plotting symbols
represent different \sn-fields, and the solid line shows the best
fitted colour term for the sample. The RMS around the best fitted
line is 0.05~mag. The reduced $\chi^2$ for the fits were
$\chi^2/dof=1.06$ for $g'$ and $\chi^2/dof=1.13$ for $r'$.
% We did not manage to find any explanation for the high $r'$ $\chi^2$,
We chose to scale the error bars for each individual fit to match
$\chi^2/dof=1.0$. We add an additional $0.01$~mag in quadrature to
the fitted zero point uncertainties in order to account for the
uncertainty of the \sdss zero points. In these fits we observe a
correlation between the zero points and the colour terms of
$\rho\sim-0.6$.
%  , and $\rho\sim0.3--0.4$
%  between the individual zero points.

%
%
\begin{table}[tbp]
  \centering
  \caption{Simultaneously fitted zero points.
    \label{tb:simuzpfit}}
  \begin{tabular}{l c r c r}
    \hline
    \multicolumn{1}{c}{\sn} & $g'$ &
    \multicolumn{1}{c}{$n_g$} & $r'$ &
    \multicolumn{1}{c}{$n_r$}\\
    \hline
    \sndr & $31.78\pm0.01$ &  85 & $31.63\pm0.01$ & 28\\
    \sndu & $31.98\pm0.01$ &  51 & $32.77\pm0.01$ & 27\\
    \sndv & $32.98\pm0.01$ &  49 & $32.76\pm0.01$ & 59\\
    \sndx & $31.95\pm0.01$ &  26 & $32.79\pm0.01$ & 35\\
    \sndy & $31.80\pm0.01$ &  47 & $33.18\pm0.01$ & 43\\
    \hline
  \end{tabular}
\end{table}

% Measured Colour Terms
%
\begin{table}[tbp]
  \centering 
  \caption{Measured colour terms for \inttel, \nottel, and
    \jkt.
%    No evidence for airmass dependent second order colour terms has been
%    seen for these measurements.
    \label{tb:colourterms}}
  \begin{tabular}{l r r}%
    \hline
    \multicolumn{1}{c}{Tel.} & 
    \multicolumn{1}{c}{$g'$} &
    \multicolumn{1}{c}{$r'$}\\
    \hline
    \inttel & $0.15\pm0.01$ & $-0.01\pm0.01$\\
    \nottel & $0.11\pm0.02$ & $0.02\pm0.01$\\
    \jkt    & $0.15\pm0.02$ & $0.02\pm0.01$\\
    \hline
  \end{tabular}
\end{table}

Note that the zero point is in this case the sum of three components;
the instrumental zero point, the aperture correction to the \psf
normalisation radius and the atmospheric extinction for the given
airmass. We are also assuming that the atmospheric extinction is
\emph{colour independent}, an assumption that was further justified by
studying the residuals for the individual image catalogues. We
examined these residuals carefully for any brightness, colour, or
position dependent correlations. Minor trends with \ccd position could
be detected at the edges of the detector, but these did not affect the
overall calibration.

The fitted zero point solutions derived above can then be used
together with colour information to obtain the standard magnitudes of
other objects in the field. However, for \sn-photometry the colour
term corrections derived from stellar photometry is not applicable.
This is due to the significant differences between \sn and stellar
spectral energy distributions (\sed), and will lead to systematic
errors (\cite{2002AJ....124.2100S}, \cite{2003AJ....125..166K}). For
\sn-photometry, it is therefore often more reliable to calculate the
\sed-dependent correction, $S$-correction, using synthetic photometry
although this requires good knowledge of both the filter responses,
the \sed, and the zero point of the instrumental filter system.

\begin{figure}[!htb]
  \centering
  \resizebox{\hsize}{!}{\includegraphics{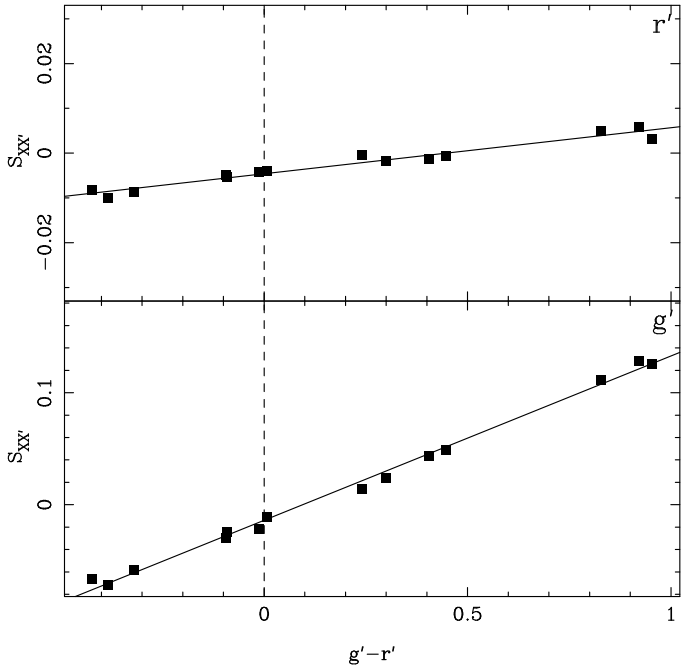}}
  \caption{Fitted linear relations for the synthetically calculated
    $S$-corrections in the $r'$ (\emph{upper} panel) and $g'$
    (\emph{lower} panel) bands.
    \label{fig:scorrection}}
\end{figure}

The calibrated magnitude in any filter system, $X'$, can be expressed
as $m_{X'} = -2.5\log_{10}(f) + ZP'$, where $f$ is the measured flux
and $ZP'$ is the zero point for that system.
The \ab magnitude in the filter system $X$ can then be expressed as
$m_{X} = m_{X'} + S_{XX'}$, where $S_{XX'}$,
\begin{equation}
  S_{XX'} =
  -2.5\log_{10}\left(%
    \frac{\int \lambda f_\lambda X_\lambda\, d\lambda}{%
      \int \lambda f_\lambda X'_\lambda\, d\lambda}\right)
  +2.5\log_{10}\left(%
    \frac{\int (X_\lambda/\lambda)\, d\lambda}{%
      \int (X'_\lambda/\lambda)\, d\lambda}\right)\, ,
  \label{eq:scorrection}
\end{equation}
is the above mentioned $S$-correction, $f_\lambda$ is the object \sed
and $X_\lambda$ and $X'_\lambda$ are the filter responses of the two
systems respectively. In equation~\eqref{eq:sdsscalibration} we
approximated $S_{XX'}$ as
\begin{displaymath}
  S_{XX'} \approx b + c_X\cdot(g_j-r_j)^\mathrm{\sdss}\,,
\end{displaymath}
dropping higher order terms. The zero point of the \inttel system,
$ZP'$ can then be expressed as $ZP' = zp - b$, where $zp$ was fitted
above, and $b$ remains to be determined. This constant does depend on
the filter responses, but it is independent of the object \sed.
Therefore we could fit it by taking the $g'$ and $r'$ magnitudes for
14 photometric standard stars (\cite{2002AJ....123.2121S}) with
colours $g'-r'<1.0$ that also have spectro-photometry
(\cite{2005PASP..117..810S}) allowing calculation of $S_{XX'}$ using
equation~\eqref{eq:scorrection}. The result of these fits are shown in
Table~\ref{tb:scorrection} and Figure~\ref{fig:scorrection}.

% Synthetic Colour terms and S-correction constant
%
\begin{table}[tbp]
  \centering 
  \caption{Fitted colour terms and $S$-correction filter
    offsets for \inttel.
    \label{tb:scorrection}}
  \begin{tabular}{c r r}%
    \hline
    Filter & \multicolumn{1}{c}{$b$} & \multicolumn{1}{c}{$c_X$}\\
    \hline
    $g'$ & $-0.014\pm0.001$ & $0.147\pm0.003$\\
    $r'$ & $-0.005\pm0.001$ & $0.010\pm0.001$\\
    \hline
  \end{tabular}
\end{table}

From the simultaneously fitted zero points in
equation~\eqref{eq:sdsscalibration} together with the filter offsets
in the first column of Table~\ref{tb:scorrection}, the zero points in
the \inttel system, $ZP'$ can then be calculated for each photometric
reference. These are presented in Table~\ref{tb:photometry}.

\subsection{Instrument Wavelength Response}
Since no \sn-free reference images were obtained with neither \nottel
nor \jkt, the \inttel references had to be used for constraining the
host galaxy models for these images as well. Differences in the
combined filter and instrument wavelength response between different
instruments can be expected and are to some extent included in
the uncertainties of the fitted kernels. However, it cannot be excluded
that the fitted \sn-fluxes could be biased when data from different
instruments are brought into the fit. Calculating the impact of this
potential systematic effect is difficult,
but a rough estimate can be obtained by studying the similarities of
the effective filter response between the different instruments. In
order to do this we repeated the recipe from the calibration, using
equation~\eqref{eq:sdsscalibration} together with \jkttel and \nottel
data and fitted the colour terms for these filters. The results are
presented along with the \inttel values in Table~\ref{tb:colourterms}.

Fortunately the colour terms for the three instruments are similar
which suggests that the systematic error for this particular case can
be expected to be small. We assumed a systematic uncertainty of
0.015~mag for all \jkt and \nottel points to account for it.

We further investigated the impact of the \jkttel and \nottel data on
the global fit by comparing the results when the whole data set was
used with light curves exclusively built from \inttel data. The
results from these two sets did not differ and therefore it can be
concluded that the \jkttel and \nottel data did not have any
systematic effect on the host galaxy model fit.

Since the calibration was tied to the \inttel system we consistently
chose to use the \inttel \wfc effective filter response functions,
provided by the Isaac Newton Group, for light curve fitting and
$K$-corrections. Any wavelength dependent difference between these
response function and the actual effective response of the system
could introduce systematics into the analysis. However, since the
synthetic calculations of the effective colour terms presented in
Table~\ref{tb:scorrection} are in good agreement with the
corresponding measured values in Table~\ref{tb:colourterms} it can be
concluded that any systematics introduced by possible incorrect
response functions are negligible.

%
%________________________________________________________________

\section{Data Analysis}\label{sec:analysis}

\subsection{Light Curve Fitting}\label{sec:lightcurve}
When \snia data is used for measuring cosmology, different correlated
properties of the \sn light curve is used in combination as a standard
candle. Intrinsically brighter \snia are typically bluer and have a
slower decline rate than intrinsically fainter objects
(\cite{1993ApJ...413L.105P}). Several different methods of combining
the measured data to a single distance dependent property have been
suggested (\cite{2001ApJ...558..359G,
2003ApJ...590..944W,2005A&A...443..781G, 2007ApJ...659..122J}).

For this work we chose the \saltii method (\cite{2007astro.ph..1828G})
to fit the light curves. In summary, \saltii consists of three
components; a model of the time dependent average \snia \sed, light
curve shape variation from the average, and a wavelength dependent
function that warps the \sed. Variation from the average \sed is
parametrised by a single light curve shape parameter, $x_1$, chosen so
that $x_1=0$ corresponds to the average decline rate, and $x_1<0$ and
$x_1>0$ represents fast and slow decline rates respectively. The
deviation, $c$, from the mean \snia $B-V$ colour at the time of
$B$-band maximum is used to parameterise the contribution of the
warping function. These two parameters are fitted for each \sn
together with the overall flux normalisation of the light curve $x_0$,
which is related to the peak $B$-band magnitude, and the time of the
$B$-band maximum. The fit is carried out in the observer
\emph{instrumental} frame rather than the standard rest frame.
$K$-corrections are therefore built in to the fitting procedure which
is a reasonable approach since they depend on the \sed of the \sn.

The model itself has been derived using both photometric and
spectroscopic data from mainly the \snls project, and is particularly
well suited for the data set presented here since high redshift
spectroscopy of the rest frame $U$-band has been used to ''train'' the
model.

The light curve data is corrected for Milky Way extinction using the
Schlegel~\etal~(1998) dust maps and the extinction law from
Cardelli~\etal~(1989) with $R_V=3.1$. Also the covariance
matrices obtained from the simultaneous $\chi^2$-fit of the \sn-fluxes
are taken into consideration for the light curve fitting after the
uncertainties of the fitted zero points are added as correlated
errors. During the fit \saltii can optionally update the weight
matrices to take uncertainties of the model and cross filter
$K$-corrections into account. In Table~\ref{tb:saltresultske} we quote
the fit results when both of these error sources were
considered.\footnote{We concluded that omitting them from the fit
  generally gives a worse goodness-of-fit, but similar fit results}
The lightcurve shape, $x_1$, and colour, $c$, corrected peak $B$-band
magnitude, $m^\mathrm{eff}_B$, is obtained as $m^\mathrm{eff}_B=m_B +
\alpha_x\cdot x_1 - \beta\cdot c$, where the values
$\alpha_x=0.13\pm0.01$ and $\beta=1.77\pm0.16$
(\cite{2007astro.ph..1828G}) are used.

\begin{figure*}[!tb]
  \centering
% \resizebox{\hsize}{!}{\includegraphics{figures/lightcurves.eps}}
  \includegraphics{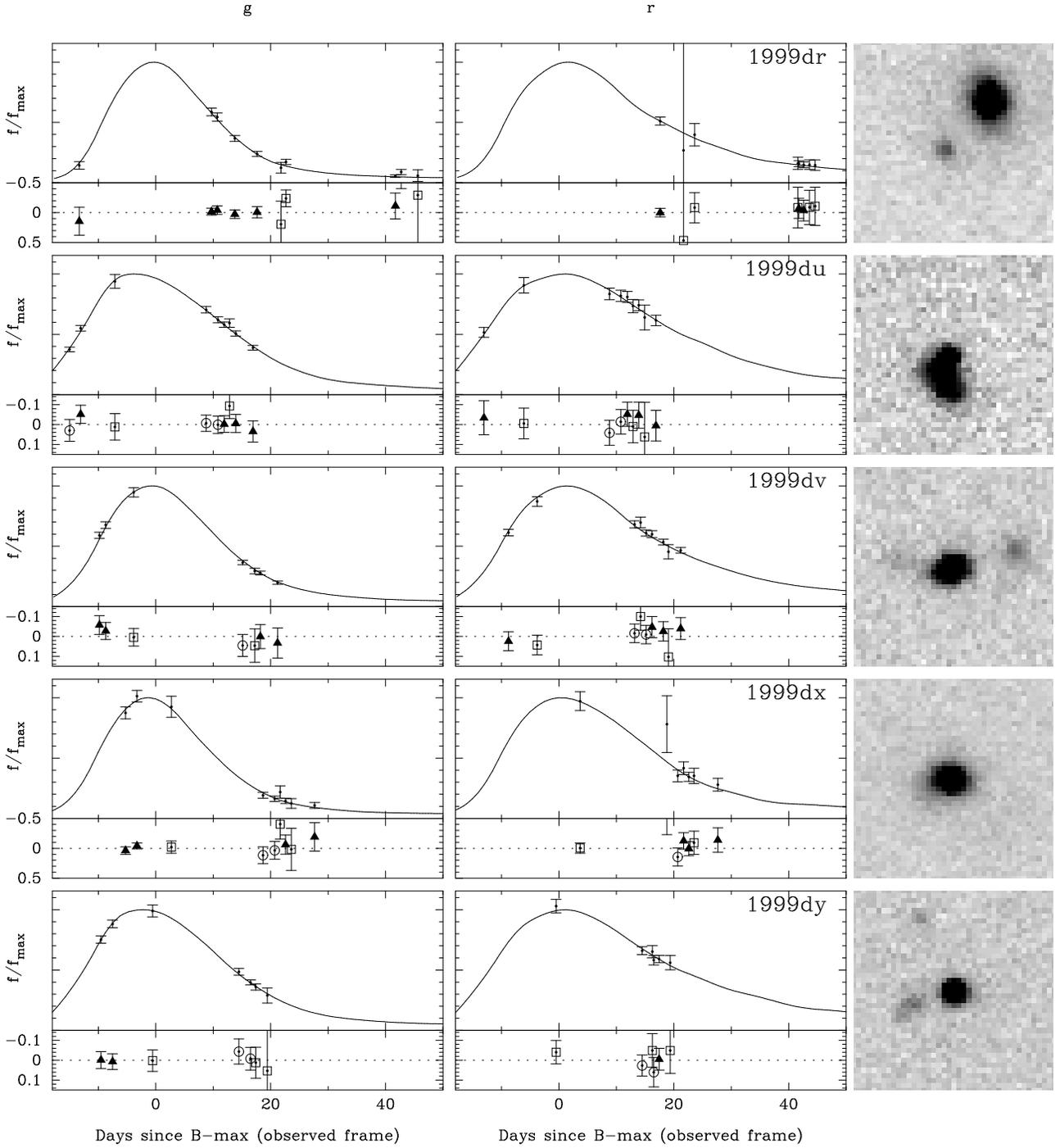}
  \caption{Light curve fits in $g'$ (\emph{left} column) and $r'$
    (\emph{middle} column), showing normalised flux against the number
    of days, in the observer frame, since the time of fitted $B$-band
    maximum. The residual plots show the \emph{magnitude} offset
    between the fitted light curve and the data from \inttel
    (\emph{triangle}), \nottel (\emph{circle}) and \jkt
    (\emph{square}). The \emph{right} column shows \inttel patches
    (13.2''$\times$13.2'') in the $g'$ band centred on the \sn.%
    \label{fig:intzlcs}}
\end{figure*}

\begin{figure*}[!htb]
  \centering
  \resizebox{\hsize}{!}{\includegraphics{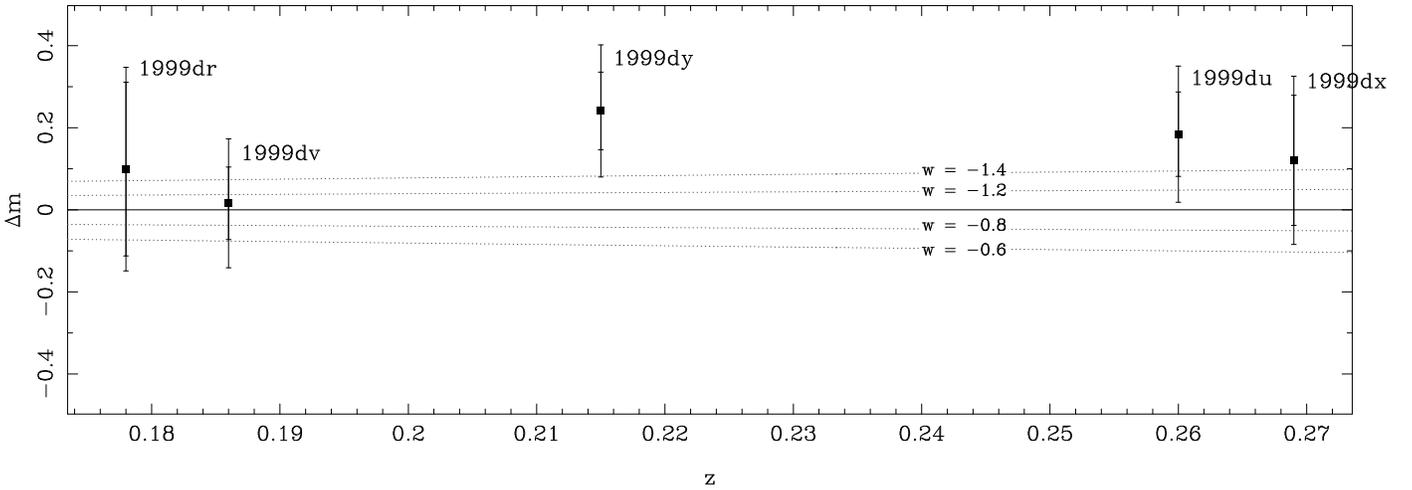}}
  \caption{Residuals from the Hubble diagram where a
    $(\Omega_M,\Omega_\Lambda,w,M_B(h_{70}))=(0.26,0.74,-1,-19.31)$
    cosmology (\cite{2006A&A...447...31A}) has been subtracted from
    the third column ($m_B^\mathrm{eff}$) of
    Table~\ref{tb:saltresultske}. The outer error bars were obtained
    by adding an assumed intrinsic uncertainty of 0.13~mag in
    quadrature to the tabulated uncertainties. The dotted lines show
    the the predicted Hubble diagrams for different values of $w$.
    \label{fig:hubble}}
\end{figure*}

The best fitted light curves are shown together with the data in
Figure~\ref{fig:intzlcs}. The stretch-colour corrected peak
magnitudes, $m^\mathrm{eff}_B$, are further compared to the
corresponding predictions for a $\Lambda$CDM universe in
Figure~\ref{fig:hubble}. Predictions for dark energy models with
an equation state parameter, $w$, differing from $w=-1$ have been
plotted as well.
% Light curves
%

% SALT2 FIT RESULTS FOR ALL SNE
%
%  - WHERE K-CORRECTION AND DIAGONAL MODEL ERRORS HAVE BEEN CONSIDERED
%
\begin{table*}[!htb]
  \centering
  \caption{Light curve fit results using \saltii.  The light curve stretch
    is calculated from $x_1$ using the relation derived in
    Guy~\etal~(2007).
    \label{tb:saltresultske}}
  \begin{tabular}{l r r r r r r r r r}
    \hline
    \multicolumn{1}{c}{\sn} &
    \multicolumn{1}{c}{$\mathrm{MJD}_\mathrm{max}$} &
    \multicolumn{1}{c}{$m^{\mathrm{eff}}_B$} &
    \multicolumn{1}{c}{$m_B$} &
    \multicolumn{1}{c}{$c$} &
    \multicolumn{1}{c}{$x_1$} &
    \multicolumn{1}{c}{$s$} &
    \multicolumn{1}{c}{$\chi^2/dof$} &
    \multicolumn{1}{c}{$dof$}\\
    \hline
    \sndr & $51413.28\pm0.76$ & $20.48\pm0.21$ & $21.59\pm0.12$ & $0.49\pm0.09$ & $-1.862\pm0.450$ & $0.826\pm0.038$ & $0.47$ & $14$\\
\sndu & $51444.15\pm0.17$ & $21.49\pm0.10$ & $21.30\pm0.04$ & $-0.03\pm0.05$ & $1.102\pm0.295$ & $1.083\pm0.028$ & $0.90$ & $14$\\
\sndv & $51439.89\pm0.11$ & $20.50\pm0.09$ & $20.79\pm0.03$ & $0.10\pm0.04$ & $-0.773\pm0.188$ & $0.912\pm0.017$ & $1.39$ & $12$\\
\sndx & $51434.45\pm0.79$ & $21.51\pm0.16$ & $21.81\pm0.05$ & $0.06\pm0.08$ & $-1.511\pm0.513$ & $0.852\pm0.044$ & $0.79$ & $12$\\
\sndy & $51438.72\pm0.21$ & $21.08\pm0.09$ & $20.94\pm0.03$ & $-0.02\pm0.05$ & $0.743\pm0.264$ & $1.049\pm0.025$ & $0.57$ & $9$\\

    \hline
  \end{tabular}
\end{table*}

% Residuals in the Hubble diagram
%

%________________________________________________________________

\subsubsection{Colour correction uncertainties}
Throughout this work, we have adopted the SALT2 technique for light curve
shape and colour corrections. This involves using an empirical colour
correction factor, $\beta$, which if associated with reddening by dust 
along the line of sight, corresponds to the total selective extinction
coefficient $\beta = R_V + 1$. However, the best fit value, $\beta=1.77\pm0.16$
(\cite{2007astro.ph..1828G}) is inconsistent with Milky Way dust properties 
($R_V=3.1$). Thus, either the measured reddening indicates dust properties in
the \sn hosts which are quite different from Galactic dust, or that
the measured colours are dominated by intrinsic \sn variations,
\eg Nobili \& Goobar (2008) find an average of $R_V=1.75 \pm 0.27$ for
a sample of near-by \sne but only $R_V \sim 1$ for a low reddening
subset of the data

The assumptions on reddening law or circumstellar dust are particular
important for one of the \sne, \sndr. The implications are further
discussed in the next section.
 
%
%________________________________________________________________

\section{Discussion}\label{sec:discussion}
The five intermediate-z \sne in our sample are well fitted by SALT2,
as shown in Figure~\ref{fig:intzlcs} and Table~\ref{tb:saltresultske}
for different \sne when the model errors are introduced. These appear
to account for the light curve shape diversities that remain after
stretch and colour warp has been considered. The average reduced
$\chi^2$ of the sample is $\sim0.81$.

The fitted light curve parameters all fall within the ''normal'' \snia
range, with the exception of the colour
of \sndr, that suggests significant reddening. An extinction of this
magnitude is unexpected both since the \sn is located in the outer
rim of the presumed host galaxy%
\footnote{The distance between the \sn and the core of the host can be
  determined to be $4.3\alpha$ after an exponential function,
  $\exp(r/\alpha)$ (where $r'$ is the distance from the core), is
  fitted to the galaxy profile. In physical units, the separation
  between \sn and galaxy core is $4.3''$, which for $z=0.183$ and a
  $\Lambda$CDM universe, corresponds to $\sim12.8\,$kpc.} and since
  the host spectrum (\cite{2006A&A...445..387B}) does not show any
  signs of star formation (see Table~\ref{tb:hostclass}) which is
  usually associated with a dusty environment.
\begin{table}[htb]
  \centering
  \caption{Quoted host galaxy types from our spectroscopic
    classification in Balland~\etal~(2006).\label{tb:hostclass}}
  \begin{tabular}{c l l}
    \hline
    \multicolumn{1}{c}{\sn} &
    \multicolumn{1}{c}{Host Type} &
    \multicolumn{1}{c}{Comment}\\
    \hline
    \sndr & Sa/Sb        & No emission\\
    \sndu & Sc/Starburst & Clear ident. Sc\\
    \sndv & Sc/Starburst & Poor gal. signal\\
    \sndx & E/S0         & Emission lines\\
    \sndy & Sc/Starburst & Faint\\
    \hline
  \end{tabular}
\end{table}
However, it should be pointed out that we lack data around maximum
for this \sn and the first $r'$-band data point was obtained as late
as $\sim17$ days after $B$-band maximum. Estimating peak properties
based on extrapolations of the model from early and late time data
should always be taken with a grain of salt. This is emphasised even
further if the first $g'$-band point is excluded from the \sndr light
curve fit. This has very little impact on the fitted stretch, that
seems to be well constrained by the post-max points. However, the
fitted peak $B$-band magnitude drops with $0.25$~mag and the colour
with $0.14$~mag (the effect on the colour corrected peak magnitude
is however small). This shift is at the $1\sigma$ level when the
errors of the fit omitting the first $g'$-band point are considered.

We also would like to point out that if the $r'$-band data is omitted
completely, and the light curve properties are fitted using only the
better sampled $g'$, the estimated value of $m_\mathrm{eff}^B$
(assuming $c=0$) deviates from a $\Lambda$CDM universe at the
$7\sigma$ level.

Two other \sne, \sndv and \sndx, show signs of potential reddening.
\sn\sndv is located very close to the core of what appears to be a
star forming galaxy based on the \sn+host spectrum, and the amount of
extinction, or intrinsic colour variation for that matter
(\cite{2003A&A...404..901N,arXiv:0712.1155}), in this case is not
unexpected. \sn\sndx appears to be hosted by an early-type galaxy, but
the spectrum does reveal emission lines ([OII]), suggesting ongoing
star formation. The confirmation spectroscopy of the remaining \sne in
the sample indicates that \sndu and \sndy are hosted by star forming
galaxies. These two \sne do not show any signs of being reddened.

After stretch and colour correcting the peak $B$-band magnitudes, the
sample (including \sndr) is consistent with the
$(\Omega_M,\Omega_\Lambda,w,M_B(h_{70}))=(0.26,0.74,-1,-19.31)$
cosmology measured by Astier \etal~(2006), giving a total $\chi^2$ of
$\chi^2\sim4$ where an intrinsic dispersion of 0.13~mag has been
assumed. Despite the good agreement in terms of $\chi^2$ it is notable
that points are generally fainter than what is predicted by the model
with a mean residual of $0.12$~mag. The probability of this happening,
assuming that all five points are drawn from the same distribution, is
$\sim15\,\%$, \ie the effect is less than $\sim1.5\sigma$. It is
possible that this deviation may originate from the rest frame
$U$-band contribution to the $g'$-band flux at these redshifts. Our
limited understanding of the \snia spectrum at these wavelengths could
possibly bias the $K$-corrections and the estimated peak $B$-band
brightness. The fact that the two bluest \sne of the sample are also
the ones that land the furthest away from the Hubble line is
consistent with this hypothesis. Data from the \sdss \sn-survey, soon
to be released, will be essential to further investigate the
uncertainties associated with the UV-part of the spectrum and for
probing this region of the redshift space.

It should be pointed out that the fitted dates of $B$-band maximum
presented here differ from the ones presented in the second column of
Table~5 in Balland \etal~(2006) but those values were based on both
preliminary photometry and calibration (as well as a different light
curve fitting procedure).

\section{Conclusions}
We have presented $g'$ and $r'$-band light curves of five \sne\ at
intermediate redshift $z\sim0.2$ obtained with the \inttel, \nottel
and \jkt telescopes. The photometry was extracted by simultaneously
fitting the \sn flux and the background galaxy. Four \sne were
discovered and have data in at least one filter before rest frame
$B$-band maximum. This constrains the fitted light curve and the light
curve shape and colour corrected peak $B$-band magnitude could be
estimated to $\sim10\,\%$.

One \sn of our sample (\sndr) shows an unusual red colour, given that
the \sn is located far away from the core of its early-type host
galaxy. Unfortunately the light curve sampling around its peak
brightness is too poor to draw any extensive conclusions about the
intrinsic colour properties of this object.

After stretch and colour correcting the peak $B$-band magnitudes, the
data is consistent with the $\Lambda$CDM cosmology at the
$\sim1.5\sigma$ level.

\begin{acknowledgement}
  The authors would like to thank David Rubin at UC Berkeley for
  assisting in parts of the analysis.
  Ariel Goobar and Vallery Stanishev would like to thank the
  G\"{o}ran Gustafsson Foundation for financial support.
\end{acknowledgement}

\begin{appendix}
\section{A Sanity Check of the Analysis Procedure}
A reliability test of the light curve building technique was carried
out by creating a set of simulated images with known properties. In
order to mimic a real situation as much as possible, a series of
images from the \inttel \wfs observations of \sndy in the $g'$-band,
were used as a template for this exercise. From this data set, the
dates, exposure time, zero points and sky background values were
borrowed.

First, the robustness and accuracy of the \psf photo\-metry was
tested. It is essential that the \psf photometry can be trusted over a
broad magnitude range, and that the field stars can be used for
calibrating the \sn light curve. For this purpose, 500 stars with a
uniform magnitude distribution between $16<m<24$, were simulated and
added to the image set. The stars were randomly positioned across the
chip, and a Moffat function (\cite{1969A&A.....3..455M}) was used as
the basis for the \psf shape, but was allowed to vary between
different images to mimic different observing conditions. Finally,
the appropriate sky background was added to each individual image, and
shot-noise was simulated.

A brightness catalogue was built for each image using the technique
described in \secref{sec:photometry}, and $15\,\%$ of the objects in
the catalogue were used to fit the zero point of the image by
comparing their measured magnitudes with the input mock catalogue. The
zero point was studied over the full magnitude range in bins of
$0.4$~mag to make sure that it was consistent and did not vary with
brightness.

The light curve building was tested by creating a second set of images
with 250 field stars, using the same magnitude distribution as above.
Additionally, 50 galaxies, all hosting \sne, were added and
where all \sne shared the same magnitude in order to later simplify
the comparison. The galaxies were modelled with elliptical Gaussian
functions and the \sn position was chosen randomly $<1.5\sigma$ from
the galaxy core. The sizes of the galaxies were allowed to float
within ~$20$\ts\% and the galaxy brightness within one magnitude from
the \sn-brightness.

The images were processed using the procedure described above and
light curves were built for all 50 \sne, by both considering the
images individually, and by first stacking images with identical
observing dates. The recipe was repeated by trying different simulated
observing conditions and \sn magnitudes, and no discrepancies between
the fitted \sn-fluxes and the mock values exceeding the statistical
expectations were detected. The correlation in magnitude between
different epochs was measured to $0.2<\rho<0.5$, depending on the
separation between \sn and host and their brightness ratio.
\end{appendix}

\Online

% 
%________________________________________________________________

\end{document}